\documentclass[preprint,flushrt]{aastex}

\newcommand{\tnstar}{\tablenotemark{*}}
\newcommand{\tndstar}{\tablenotemark{**}}
\newcommand{\tndag}{\tablenotemark{\dag}}
\newcommand{\tnddag}{\tablenotemark{\dag\dag}}

\begin{document}

\title{Very Isolated Early-Type Galaxies}
\author{John T. Stocke and Brian A. Keeney}
\affil{Center for Astrophysics and Space Astronomy, Department of Astrophysical and Planetary Sciences, Box 389, University of Colorado, Boulder, CO 80309}
\author {Aaron D. Lewis}
\affil{Dept. of Physics \& Astronomy, 4171 Reines Hall, UC, Irvine, CA 92697}
\author{Harland W. Epps}
\affil{UCO/Lick Observatory, Natural Science 2, UC, Santa Cruz, CA 95064}
\author{Rudolph E. Schild}
\affil{Center for Astrophysics, 60 Garden St., Cambridge, MA 02138}
%email{stocke@casa.colorado.edu, brian.keeney@colorado.edu, lewisa@uci.edu}
\shorttitle{Isolated Elliptical Galaxies}
\shortauthors{Stocke et al.}

\begin{abstract}
We use the Karachentseva (1973) ``Catalogue of Very Isolated Galaxies'' to investigate a candidate list of $>\,$100 very isolated early-type galaxies. Broad-band imaging and low resolution spectroscopy are available for a large fraction of these candidates and result in a sample of 102 very isolated early-type galaxies, including 65 ellipticals and 37 S0 galaxies. Many of these systems are quite luminous and the resulting optical luminosity functions of the Es and early-types (E+S0s) show no statistical differences when compared to luminosity functions dominated by group and cluster galaxies. However, whereas S0s outnumber Es 4:1 in the CfA survey, isolated Es outnumber S0s by nearly 2:1. We conclude that very isolated elliptical galaxies show no evidence for a different formation and/or evolution process compared to Es formed in groups or clusters, but that most S0s are formed by a mechanism (e.g., gas stripping) that occurs only in groups and rich clusters. Our luminosity function results for ellipticals are consistent with very isolated ellipticals being formed by merger events, in which no companions remain.
 
CHANDRA observations were proposed to test specifically the merger hypothesis for isolated ellipticals. However, this program has resulted in the observation of only one isolated early-type galaxy, the S0 KIG 284, which was not detected at a limit well below that expected for a remnant group of galaxies. Therefore, the hypothesis remains untested that very isolated elliptical galaxies are the remains of a compact group of galaxies which completely merged.
\end{abstract}

\keywords{galaxies: elliptical and lenticular, cD --- galaxies: individual (KIG 284) --- galaxies: luminosity function, mass function --- X-rays: galaxies}

\section{Introduction}

Because the two-point correlation function of galaxies is very steep \citep[e.g.,][]{peebles93}, the best place to find a galaxy is next to another galaxy. Another way of saying this is that truly isolated galaxies are exceptionally rare in the Universe. For example, \citet{tully87} finds no completely isolated galaxies at all within the local supercluster; all are members of small or large bound groups or loose associations. And yet the mythical ``field population'' of galaxies continues to be referenced in the literature as a comparison to various cluster, group, compact group and interacting galaxy studies \citep[e.g.,][]{zabludoff96,koopmann98,toledo99,christlein03}. Still, it should be possible to locate some galaxies which are very isolated relative to the much larger number which are members of clusters, small bound groups or loose and still unbound associations. If the case can be made that a potentially isolated galaxy has not experienced a merger or interaction with another galaxy for a time much longer than the timescale for the physical process under study (e.g., 10$^8$ yrs for a starburst; 10$^7$ yrs for an AGN active phase; few $\times 10^7$ yrs for spiral galaxy density wave generation), then these very isolated galaxies provide an excellent baseline comparison sample for that property. For example, \citet{haynes80,haynes84a} and \citet{haynes84c} used an isolated galaxy sample as a baseline for the \ion{H}{1} properties of galaxies; \citet{adams80} have shown that very isolated galaxies are deficient in radio continuum emission compared to other galaxies; and \citet{koopmann98} have shown that spirals in the Virgo cluster have different structural properties than more isolated spirals. In all of these cases, the use of an isolated galaxy comparison sample allows us to infer something about the effects that an external environment can have on the internal characteristics of a galaxy. For this reason, a large sample of very isolated galaxies is extremely valuable, providing an important comparison sample, which facilitates studies of environmental effects on galaxies.
 
Based upon a visual inspection of all $\sim 30,000$ bright (m$_B \leq 15.7$) galaxies catalogued by \citet{zwicky57}, \citet{karachentseva73} listed over 1000 Zwicky galaxies that are very isolated. Because few redshifts of these galaxies were available at the time of Karachentseva's work, she based her isolation criterion on the observed angular sizes and distances between galaxies. Thus, this sample is representative, not complete, because other isolated galaxies would not be included if they were projected onto foreground or background galaxies of comparable angular size (see below). Surprisingly, Karachentseva's (1973) list contains over 100 galaxies which she classified morphologically as early-types (Es and S0s).

The possible existence of very isolated early-type galaxies is unexpected based upon the typical environment for such systems \citep[e.g.,][on the morphology/density relationship]{dressler84,oemler92}. And yet recent theoretical \citep[e.g.,][]{barnes92,barnes85,athanassoula97} and observational work \citep[e.g.,][]{mulchaey98,mulchaey99,zabludoff03} suggest that very isolated elliptical galaxies could be the final outcome in the evolution of dense groups of galaxies that have completely merged, leaving only a single, large elliptical galaxy behind. The detection of the isolated elliptical NGC 1132 as a diffuse X-ray source by \citet{mulchaey99} at a luminosity (L$_x \approx 5 \times 10^{42}$h$^{-2}_{70}$ ergs s$^{-1}$) comparable to the L$_x$ of elliptical-dominated groups of galaxies is significant new evidence in favor of the merger hypothesis for forming ellipticals in general. Thus, the number, detailed structure, and X-ray properties of very isolated elliptical galaxies provide new tests of the merger hypothesis for elliptical galaxies. Are these true elliptical galaxies or do they possess systematic differences from the ellipticals found in rich groups and clusters? Similar questions could be asked about very isolated S0 galaxies, whose detailed histories are even more poorly understood. Proposed scenarios include ram-pressure stripped spirals \citep{quilis00}, mergers of large galaxies with small companions \citep{mihos95b,bekki98}, and early starbursts, followed by passive evolution \citep{welch03}. Only two of these possible histories can account for isolated S0 galaxies.
  
In this Paper, we begin the study of very isolated early-type galaxies by scrutinizing a list of candidate very isolated early-type galaxies drawn from \citet{karachentseva73}, presenting a final list of very isolated E \& S0 galaxies and computing the optical luminosity function (LF) for them. In addition, we present a single CHANDRA ACIS imaging analysis of one of the most luminous galaxies in this sample, a first attempt at testing the merger hypothesis for the formation of very isolated ellipticals by detecting diffuse X-ray emission.

In the following section we describe the \citet{karachentseva73} sample and present the sample of isolated early-type galaxies identified from it through further scrutiny. In Section 3, we present the optical luminosity functions of the ellipticals and the S0s in the sample. Section 4 presents the CHANDRA and optical imaging and spectroscopy results for the one isolated early type galaxy observed by CHANDRA: KIG 284. Section 5 includes a brief discussion and conclusions drawn from this work. Throughout this work we adopt the NASA Extragalactic Database (NED) abbreviation for the Karanchentseva (1973) galaxies (KIG) as well as H$_0 = 70$ km s$^{-1}$ Mpc$^{-1}$.

\section{The Isolated Early-type Galaxy Sample}

The early-type galaxy sample we have selected is drawn from the ``Catalogue of Very Isolated Galaxies'' by \citet[KIG Catalogue hereafter]{karachentseva73}. This catalogue contains 1051 galaxies with m$_B \leq 15.7$, chosen from the ``Catalogue of Galaxies and of Clusters of Galaxies'' \citep[CGCG hereafter;][]{zwicky57} by inspecting the Palomar Observatory Sky Survey (POSS) for isolation. Since the CGCG contains $> 30,000$ galaxies, demonstrably isolated galaxies are $\lesssim 3$\% of the total galaxy population and so are very rare. However, the CGCG is a magnitude-selected sample and not an angular size selected sample like the \citet{nilson73} Catalogue. As such, the CGCG selection method is now known to be biased against the selection of low surface brightness galaxies \citep{mcgaugh95}, and low surface brightness galaxies may be systematically more isolated than higher surface brightness galaxies \citep{mcgaugh96,impey97}. However, this selection bias is probably not important for early type galaxies, since these objects have quite high central surface brightnesses due to their luminous bulge components.
 
The KIG Catalogue selection criterion uses the POSS alone by defining an isolated galaxy to be a CGCG galaxy which has no ``companions'' within 20 galaxy diameters. Companions are defined to be galaxies within a factor of 4 in angular diameter of the isolated galaxy candidate \citep[i.e., $\sim \pm\,$2 mag in luminosity using the relationships in][]{hutchmeier87}. While the exact density of the galaxy environment is difficult to measure for most galaxies due to projection effects, an apparently isolated galaxy cannot be ``created'' by projection effects. Thus, these KIG galaxies are truly isolated. For example, based upon the above selection criteria, and assuming that the projected galaxy diameter on the sky is $\sim 20$ kpc, and that the peculiar velocity of these systems is $\sim 300$ km s$^{-1}$, a KIG galaxy has not suffered an encounter with another large galaxy for at least the past billion years (5-10 rotation periods for a large spiral galaxy). This is the case for these KIG galaxies unless the companion galaxy is much smaller or larger than the KIG candidate galaxy, or the isolated galaxy has merged recently with all of its neighbors.
 
Since it is surprising that the KIG Catalogue has 100+ galaxies which are suggested to be early-type systems (Es and S0s), the verification of this sample and its properties is important. In order to be certain that no early type galaxies were misclassified, a total of 206 KIG galaxies were examined on the POSS and, for the most part, their morphological classifications were confirmed. In the course of checking the morphologies, we also checked the isolation of these galaxies; only 13 galaxies were eliminated from consideration here owing to comparably sized companions, evidently missed by \citet{karachentseva73}. These multiple inspections \citep{adams80} found the following breakdown of morphologies: 109 Es and S0s, 56 uncertain classifications, but nonetheless not early types, and 28 spirals. However, some of these objects are faint and compact enough that these POSS-derived classifications are not persuasive in all cases. For this reason we obtained red optical images for 84 of the 109 Es and S0s at the Mt. Hopkins 0.6m telescope in the mid 1980s and aperture spectra for all of them at the Lick Observatory 3m and the Steward Observatory 2.3m during the same time period. Additionally, photographic images had been previously obtained for seven others by \citet{adams80}. The optical imaging eliminated seven Es and S0s, which are spirals with weak, but visible spiral structure. The remaining 102 early type KIG galaxies are listed in Table \ref{tbl:objlist}.

CCD imaging was obtained for 84 of these KIG galaxies at the Mt. Hopkins 0.6m telescope through an ``F'' filter \citep{schild84}, a broad-band filter whose combined throughput and CCD response mimicked the Kodak photographic F emulsions in use at the time. The effective wavelength of this filter is closely approximated by a Gunn r filter \citep{schild81}. Sixty-six of these images were sufficient to allow detailed surface brightness profiles to be extracted and fit to de Vaucouleurs r$^{1/4}$-laws. Objects well-fit by r$^{1/4}$-laws were classifed as ellipticals, while those objects showing even slight evidence for disks (at intermediate radii) were classified as S0s. The dividing line between these two classifications occurred if more than 5 radial bins were in excess of the r$^{1/4}$-law fits by $\geq 1\sigma$. The radial bins were set to the observed seeing in these images (typically 2 arcsecs) and there were typically 15-25 bins in each profile. Only seven galaxies were close to the E/S0 dividing line, with 4 being classified as Es. The remainder were either too close to bright stars and/or had sky too bright to allow an unambiguous extraction of the surface brightness profile. These were classified by eye either from the 0.6m CCD images (12 objects) or the \citet{adams80} photographic images (7 objects) and have morphologies marked with asterisks in Table \ref{tbl:objlist}. We were unable to obtain new images of the remaining 18 galaxies in Table \ref{tbl:objlist}, whose classifications were made using the POSS. The morphologies of these 18 galaxies are marked with double asterisks. The KIG early-type galaxy sample in Table \ref{tbl:objlist} has 65 ellipticals and 37 S0 galaxies. Thus, 80\% of the Es and 86\% of the S0s were classified using new images.

A substantial difference exists in the relative numbers of Es and S0s in the CfA survey volume \citep[4:1 in favor of S0s;][]{marzke94b} compared to KIG Catalogue (1.75:1 in favor of ellipticals). This could be an indication of the relative difficulty in forming Es and S0s in very sparse environments, as it is unlikely that the morphological typing could be this inconsistent. The \citet{marzke94b} typing is from \citet{devaucouleurs76} or \citet{nilson73} with a few additions from J. Huchra in conjunction with \citet{marzke94b}. All of the CfA morphological types were judged by visual inspection of available plate material while those herein are primarily based upon surface brightness profiles. Thus, some systematic differences may be present between these two samples. But while we can imagine a few differences that might have allowed some bone fide S0s to be present in the \citet{marzke94b} sample of ellipticals and for us to have misclassified some bone fide isolated Es as S0s, it is hard to imagine that these differences are so great as to create the large population difference between these two samples. Thus, we believe that these morphological differences are real and that very isolated S0 galaxies are extremely rare.

The spectroscopy of all 102 galaxies in Table \ref{tbl:objlist} was obtained using either the Lick 3m spectrograph with the image dissector scanner detector or the Steward 2.3m blue reticon spectrograph. The dispersion of these spectra was 7-8 \AA\ covering 3400-6400 \AA\ at a signal-to-noise ratio of 5-10 per resolution element. Individual emission and absorption lines were identified and redshifts obtained by measuring wavelengths of individual features. Most of these galaxy spectra showed only absorption lines typical of old stellar populations but a few had emission lines in their spectra. Where emission lines are present, they are indicated in Table \ref{tbl:objlist}. While not up to the standards of modern cross-correlation techniques, we estimate that the redshifts in Table \ref{tbl:objlist} are accurate to $\pm150$ km s$^{-1}$, based upon redshift agreement between spectral features measured. This accuracy is quite adequate for determination of the luminosity function of these galaxies.

Table \ref{tbl:objlist} gives the following basic information on the KIG E+S0 galaxy sample by column: (1) KIG number; (2) adopted morphological classification (see above); (3) apparent Zwicky magnitude (m$_{Zw}$) given in the CGCG; (4) Galactocentric recession velocities obtained from our spectra and corrected to the Galactic frame (km s$^{-1}$); (4) Absolute blue luminosity derived from the values in columns (3) and (4), corrected for Galactic extinction and assuming H$_0 = 70$ km s$^{-1}$ Mpc$^{-1}$. The extinction corrections were calculated for each galaxy using the data from Burstein \& Heiles (1982) to determine E(B--V) and assuming A$_B=4.0\,$E(B--V). This approach was used to be consistent with a similar approach used by \citet{marzke94a,marzke94b}, to which we will compare luminosity functions in the next section. Because the CGCG is restricted to higher Galactic latitudes, the extinction corrections are modest (E(B--V$) \leq 0.2$). The final column (5) lists any emission lines or Balmer absorption lines present in our optical spectra. Any other comments about the individual objects are also placed here. The complete sample of ellipticals (or S0s) used in the next Section to compute the luminosity function of Es and E+S0s are marked next to their KIG numbers with double daggers and single daggers, respectively.

\section{Luminosity Functions (LFs) of Isolated Ellipticals and S0s}

The interesting result apparent in Table \ref{tbl:objlist} is that many of these isolated Es and S0s are extremely luminous galaxies. In order to compare the absolute magnitude distribution with ellipticals found in other surveys, we have computed the optical LFs of isolated Es and all isolated early-type galaxies and compared it with the E and E+S0 LFs of \citet{marzke94a,marzke94b}, which is based upon the expanded CfA galaxy redshift survey \citep{huchra83}.

It is well-known that the CGCG Catalogue shows evidence for incompleteness at the faint end \citep[m$_{Zw} =15.5$-15.7;][]{huchra76, haynes84b}. A $<$V/V$_{max}$$>$ calculation for the entire list in Table \ref{tbl:objlist} confirms incompleteness for this sample specifically. So, we have set the limiting magnitude for luminosity function calculations at m$_{Zw} =15.4$. In addition, Table \ref{tbl:objlist} shows a relative dearth of KIG galaxies at lower redshifts. While this is partially due to the smaller volume sampled at lower recession velocities, it is likely that the sample selection criteria, specifically the isolation criterion, exclude lower recession velocity galaxies systematically. In most directions $cz \leq 3000$ km s$^{-1}$ places galaxies within the confines of the Local Supercluster, where there are few, if any, truly isolated galaxies \citep{tully87}. While there are a few KIG galaxies in Table \ref{tbl:objlist} which are at these recession velocities, the isolation selection criterion biases against their inclusion. Therefore, we have excluded the volume of the Local Supercluster from our search area as well as those few galaxies with m$_{Zw} \leq 15.4$ which have $cz \leq 3000$ km s$^{-1}$: 1 elliptical (KIG 833) and 4 S0s (KIG 89, 141, 503, \& 769). The remaining sample sizes are: 26 ellipticals and 45 early-type galaxies total. A $<$V/V$_{max}$$>$ completeness test for the 26 ellipticals (45 E+S0s) yields 0.532 $\pm$ 0.057 (0.556 $\pm$ 0.043), suggesting that these samples are complete (nearly complete). Eighty percent of the E and 86\% of the E+S0 samples used to compute the LFs were classified using new images. If all KIG galaxies in Table \ref{tbl:objlist} were included in the comparison below, it would not alter the conclusions drawn.

\begin{figure}
\begin{center}
\plotone{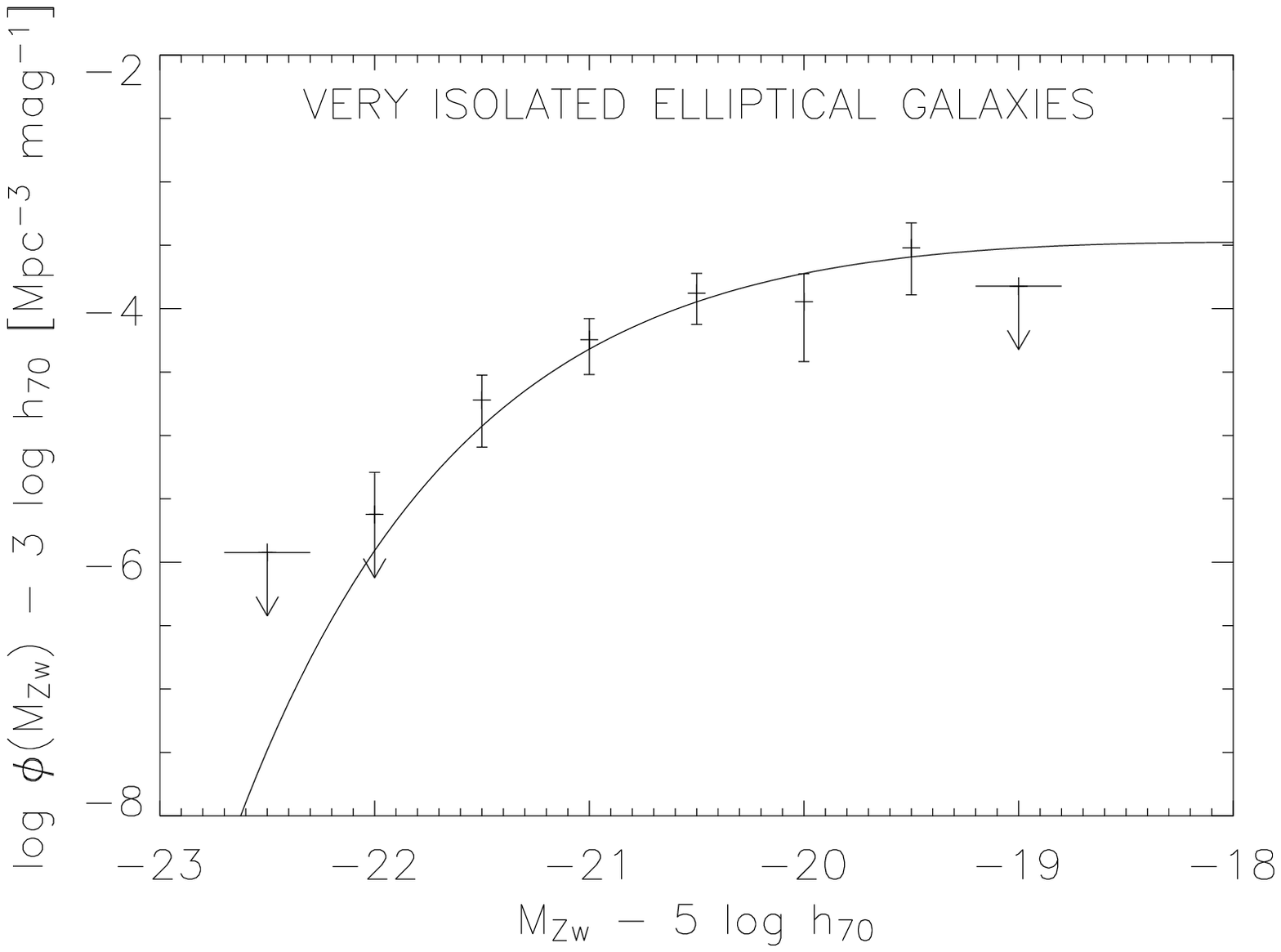}
\end{center}
\caption{Luminosity function of the elliptical galaxy sample (26 members), vertically scaled to match the \citet{marzke94b} normalization. The solid line is the Schechter function fit to the \citet{marzke94b} data. The mag bins are 1/2 mag wide and the vertical error bars indicate $\sqrt{\rm N}$ statistics for
each bin.}
\label{fig:E_LF}
\end{figure}

\begin{figure}
\begin{center}
\plotone{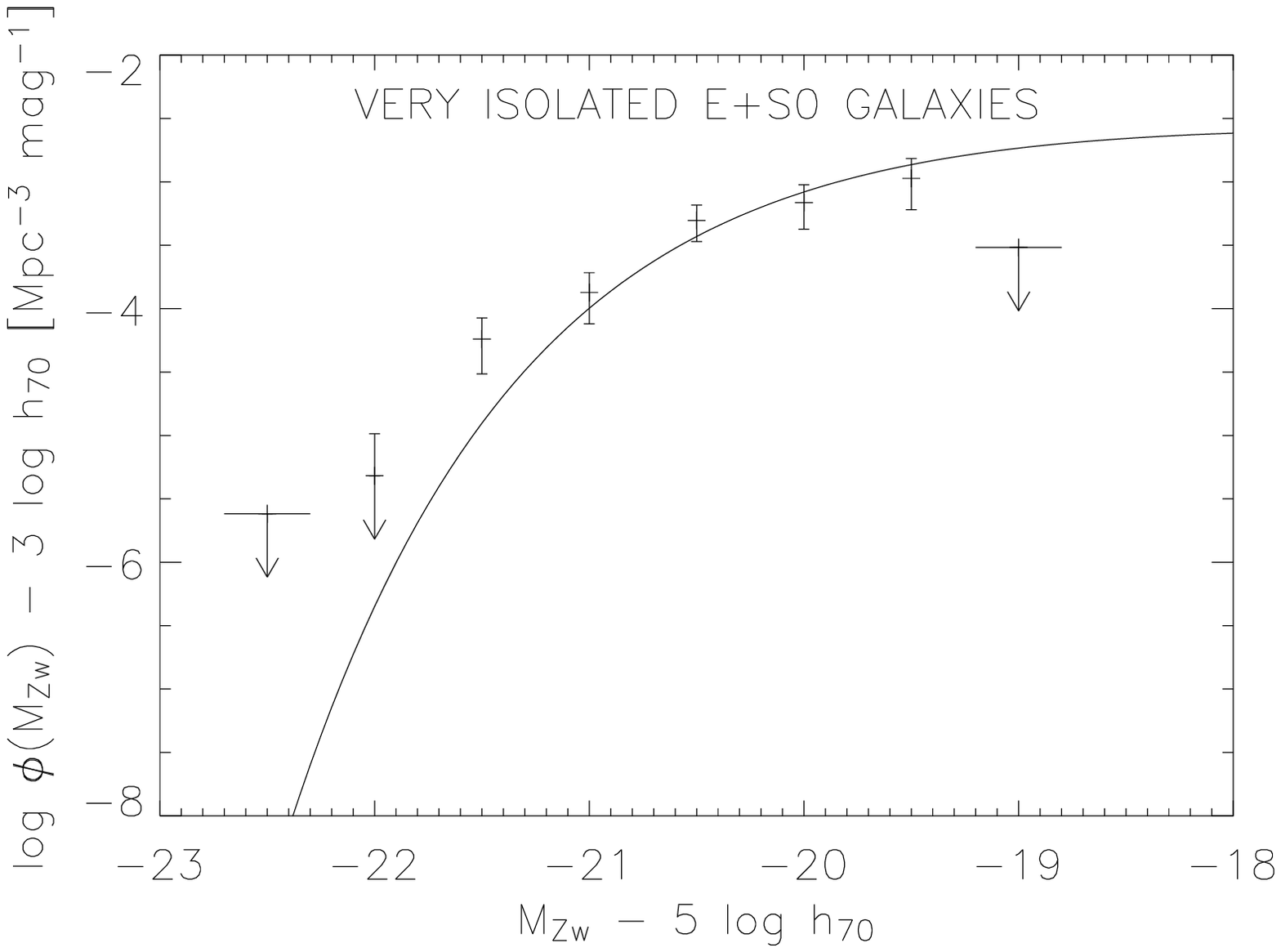}
\end{center}
\caption{Luminosity function of the E+S0 galaxy sample (45 members), vertically scaled to match the \citet{marzke94b} normalization. The solid line is the Schechter function fit to the \citet{marzke94b} data. The mag bins are 1/2 mag wide and the vertical error bars indicate $\sqrt{\rm N}$ statistics for each bin.}
\label{fig:E+S0_LF}
\end{figure}

Luminosity functions for these two samples are shown in Figures \ref{fig:E_LF} and \ref{fig:E+S0_LF} computed using the N/V method \citep{felten76} in half magnitude bins and compared to the functional fit to the elliptical and E+S0 LFs of \citet{marzke94b}. The \citet{schecter76} function parameters derived by \citet{marzke94b} from their data but converted to  H$_0 = 70$ km s$^{-1}$ Mpc$^{-1}$ are: 

Ellipticals: M$^* = -20.0$; $\alpha = -0.85$; $\Phi^* = 5.1 \times 10^{-4}$ galaxies Mpc$^{-3}$ mag$^{-1}$

E + S0s:~~~$\:$M$^* = -19.6$;  $\alpha = -0.92$; $\Phi^* = 3.4 \times 10^{-3}$ galaxies Mpc$^{-3}$ mag$^{-1}$

The isolated galaxy LF data have been scaled upwards in galaxy density by factors of $\sim 100$ and 200 to match the \citet{marzke94b} data for Es and E+S0s respectively, because we know that our selection criteria excludes most of the early-type galaxies in the survey volume. Therefore, this procedure is justified as we are testing the numbers of luminous and less luminous galaxies in the isolated sample, relative to the LF of early-type galaxies in general. No horizontal offsets are required since both our data and the \citet{marzke94b} data use Zwicky magnitudes \citep{delapparent03}. The important point of the comparisons shown in Figures \ref{fig:E_LF} and \ref{fig:E+S0_LF} is that, once scaled, the overall shapes of the LFs match pretty well. The scaling was accomplished in both cases using a $\chi^2$ fitting procedure for our data to the \citet{marzke94b} Schechter functions shown. The reduced $\chi^2$ values are 0.39 and 1.02 respectively for the Es and E+S0s. The larger $\chi^2$ value for the E+S0 sample appears to be due to a slight excess in number density of KIG S0 galaxies at M$_{Zw} \leq M^*$; i.e., while the CfA redshift survey found a less luminous M$^*$ value for E+S0s compared to Es alone, our S0 sample is predominantly more luminous than M$_{Zw} = -19.6$, thus yielding a slightly poorer (but still acceptable) fit.  If instead, we fit our E+S0 data to the elliptical LF from \citet{marzke94b}, the reduced $\chi^2$ value is 0.23. The excellent fit to the E LF parameters by our E+S0 data could be due to some small inconsistencies in the morphological typing between these two samples as described in the previous section. However, the recent concerns described by \citet{delapparent03} involving contamination of early-type LFs by dwarf galaxies of uncertain types are not relevant here since our comparisons do not extend below ${\rm M} \geq -19$.

At the high luminosity end, the KIG elliptical sample contains one galaxy (KIG 701) which is at $\approx 7$L$^*$. And the absence of even more luminous isolated ellipticals is not precluded because the sample volume is not nearly large enough to constrain their space density (note upper limits at high luminosity in Figures \ref{fig:E_LF} \& \ref{fig:E+S0_LF}). This is true either if the isolated Es obey a standard Schechter function or even if they have excess numbers above a Schechter function, as seen in rich clusters due to the cD galaxies. In other words, we cannot preclude the existence of very isolated cD-like galaxies, although it would require a much larger search volume to find one. In summary, there is no indication that the KIG sample is not drawn from the same parent population as the Schechter functions shown in Figures \ref{fig:E_LF} and \ref{fig:E+S0_LF}.

These $\chi^2$ values do not change substantially even if all the KIG galaxies in Table \ref{tbl:objlist} are included in the LF comparisons (factor of two larger $\chi^2$ values, largely due to incompleteness; see above). Thus, we conclude that there is no evidence based upon luminosities that very isolated ellipticals and S0s are significantly different from early type galaxies in clusters and dense groups.

\section{X-ray and Optical Observations of KIG 284}

To further test whether isolated early-type galaxies are similar to other early-type galaxies, CHANDRA/ACIS imaging spectroscopy was obtained for one of the most luminous galaxies in the KIG sample in Table \ref{tbl:objlist}, KIG 284. This test is based upon the recent discovery that some isolated ellipticals are surrounded by X-ray emitting gas, similar in extent and L$_x$ to that found in dense, elliptical-dominated groups of galaxies \citep{mulchaey99}. However, only some isolated Es exhibit extended X-ray emission; \citet{zabludoff03} finds that only 1 in $\sim 5$ isolated ellipticals shows extended X-rays. This may indicate that at least two different merger scenarios are possible; i.e., these isolated Es can be formed either from dense groups that already possessed an intra-group medium \citep[like the elliptical-dominated groups studied by][]{mulchaey99} or from ones which did not (the Local Group?). Therefore, several KIG galaxies would need to be observed in X-rays to test this hypothesis definitively.

CHANDRA ACIS-S observations were made of KIG 284 on 2001 March 14 for a usable integration time of 8.26 ksecs; a small amount of exposure time was removed due to high background. No obvious detections of this galaxy were made and so we are able to set only upper limits on the X-ray flux of KIG 284. At the observed redshift of $z=0.043$, 1 arcsec = 850 pc and so we obtained limits on count rates within two apertures of the following radii: 58 arcsec ($50$h$^{-1}_{70}$ kpc) and 140 arcsec ($120$h$^{-1}_{70}$ kpc). We also searched for emission in a larger aperture ($200$h$^{-1}_{70}$ kpc), but this aperture extended beyond the edge of the S3 ACIS chip onto the S2 chip, which is less sensitive; we did not detect any extended X-ray emission in this largest aperture either. The smaller and larger apertures were used to search both for emission related to the individual galaxy as well as emission that has the size expected for remnant emission from a pre-existing galaxy group. In a small X-ray survey of poor groups of galaxies, \citet{mulchaey98} found evidence for two X-ray components in the elliptical-dominated groups they detected. The first has a size (30-$60$h$^{-1}_{70}$ kpc), location, and temperature ($<\,$1 keV) suggesting an association with the interstellar medium of the central, dominant elliptical in the group. The second component found was substantially larger, extending to 150-$400$h$^{-1}_{70}$ kpc, and was hotter. \citet{mulchaey98} identified this component with the entire galaxy group. For reference, the isolated elliptical NGC 1132, mentioned in Section 1, has an observed X-ray core radius of $135$h$^{-1}_{70}$ kpc, a full extent of nearly $350$h$^{-1}_{70}$ kpc and a total X-ray luminosity of $5 \times 10^{42}$h$^{-2}_{70}$ ergs s$^{-1}$ \citep{mulchaey99}. Using only the softer portion of the ACIS-S energy band (0.5-2 keV), $3\sigma$ upper limits for the 50 and 120 kpc radii apertures were 0.06 and 0.09 counts s$^{-1}$, respectively. The smaller radius recorded a $1\sigma$ excess of counts above background, so we cannot rule out a very faint detection of this galaxy in our data. However, this possible small excess is not restricted to a small number of pixels and so is not a more definite detection of a point source. Use of the full 0.3-10 keV band increases the background count rate and so provides poorer upper limits. Since the expected temperature for a poor group of galaxies or the diffuse emission from an individual elliptical or S0 galaxy is $\sim\,$1 keV, the restricted energy band is appropriate in this case. At the luminosity distance of KIG 284 these limits correspond to $< 3 \times 10^{41}$ and $4.5 \times 10^{41}$h$^{-2}_{70}$ ergs s$^{-1}$ respectively, if the emission were smoothly distributed throughout the entire aperture. But, where detections have been made of poor groups, the X-ray emission is quite centrally concentrated. So we use the first limit as a conservative upper limit on the X-ray luminosity from the KIG 284 vicinity. This limit is over an order of magnitude below the L$_x$ of NGC 1132 and is also at least a factor of 5 less than the least luminous galaxy group detected by \citet{mulchaey98}. Therefore, we can rule out KIG 284 as a remnant group of galaxies, as \citet{mulchaey99} propose for NGC 1132. However, our X-ray upper limit is in the middle of the range (L$_x \sim 10^{40-43}$ ergs s$^{-1}$) found for luminous ellipticals by \citet{eskridge95} and is a few times higher than the total L$_x$ found for the nearby X-ray faint S0 galaxy NGC 1553 \citep{blanton01}. The possible $1\sigma$ excess count rate at the location of KIG 284 would be at approximately the same L$_x$ as NGC 1553. So, KIG 284 could be a quite normal E or S0 galaxy based upon the X-ray limits we have presented.

KIG 284 was included in our target list because its 0.6m image appeared to show a low surface brightness excess at large radii, similar to what is seen in cD galaxies (although this galaxy is not nearly as luminous as a cD galaxy)
 Additionally, our Steward 2.3m reticon spectrum of KIG 284 contains emission lines which could either be due to recent star formation or to gas heated by other low-ionization parameter processes like so-called ``cooling flows''. Balmer absorption lines indicate the presence of young stars in this galaxy. Therefore, it is an interesting, although anomalous member of this sample based upon our imaging and spectroscopy.

\begin{figure}
\begin{center}
\plotone{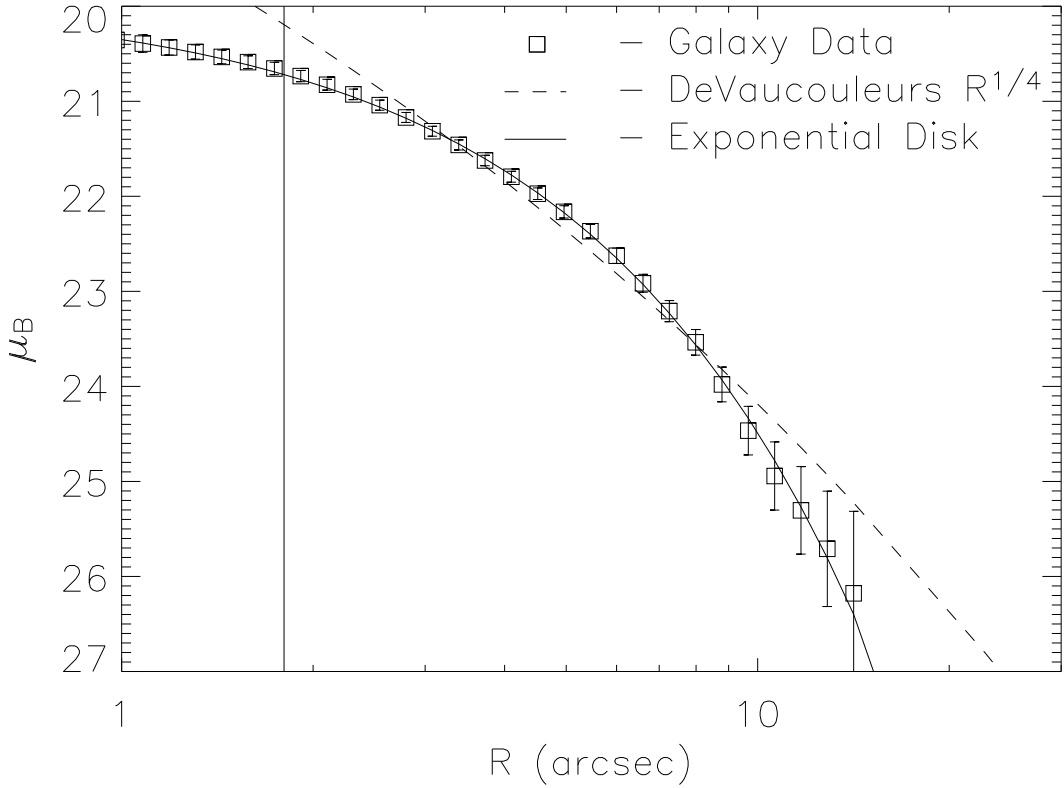}
\end{center}
\caption{Surface brightness profile of KIG 284 (squares) with best-fit de Vaucouleurs $R^{1/4}$ (dashed line) and exponential disk (solid line) models overlaid. This galaxy is less centrally concentrated than elliptical galaxies, so this profile confirms our previous classification of KIG 284 as an S0. The vertical line indicates the seeing (1.8 arcsecs) in the image from which the surface brightness distribution was derived. The small companion galaxy to the SW of KIG 284 has not been included in the surface brightness profile.}
\label{fig:KI284_surf}
\end{figure}

In order to make certain that the properties of KIG 284 were as we had originally determined from the older image and spectrum, we obtained a new spectrum using the double imaging spectrograph (DIS) and a new (B,R) image pair using the ``SPICAM'' CCD imager at the Apache Point Observatory 3.5m Telescope \footnote{The Apache Point 3.5m Telescope is operated by the Astronomical Research Consortium (ARC)}. The broad-band images were obtained on 2002 February 8 in 1.8 arcsec seeing. No surface brightness irregularities due either to dust or to weak spiral structure were visible in the B or R band images. After removing the effects of a small companion galaxy to the SW, the surface brightness profile obtained from the R-band image is shown in Figure \ref{fig:KI284_surf}. The exponential disk and r$^{1/4}$-law profiles shown in the figure are least square fits to the surface brightness data. This profile verifies our previous classification of this galaxy as an S0 and shows no evidence for an extended outer envelope that the 0.6m image appeared to show. This mistaken impression was probably due to a slight mis-determination of the sky level in our original CCD image and rules out KIG 284 as a cD-like galaxy. Using the aperture photometry package (``apphot'') within IRAF, a blue magnitude of B$ = 16.3 \pm 0.1$ and (B--R$) = 1.1 \pm 0.2$ were obtained. If we include a small companion to the southwest of KIG 284 the total magnitude increases to 16.1, which converts to m$_{Zw} = 15.5 \pm 0.4$ using the recent CCD photometry of \citet{gaztanaga00}. The much larger mag error is due to the scatter inherent to the conversion found by \citet{gaztanaga00}. Our measured galaxy magnitude is consistent with the Zwicky magnitude listed in Table \ref{tbl:objlist}.

The 3.5m DIS spectrum was obtained on 2003 January 26 and covers 3600-9700 \AA\ at $\sim\,$6 \AA\ resolution in two spectra, with somewhat lower sensitivity in the region of the dichroic (5200-5600 \AA). This spectrum confirms the presence of strong, low ionization emission lines, and adds H$\alpha$, [\ion{N}{2}] 6584 \AA\ and the [\ion{S}{2}] doublet ($\lambda\lambda$ 6717, 6731 \AA) to the list of emission lines in KIG 284 listed in Table \ref{tbl:objlist}. The emission line ratios are indicative of star formation based upon, e.g., the theoretical work of \citet{kewley01}, as are the presence of higher Balmer lines in absorption (H$\delta$, H$\epsilon$, etc; but not H$\gamma$, which is mostly filled-in by emission at our spectral resolution). Using the observed H$\alpha$ luminosity from our spectrum, we infer a current star formation rate for this galaxy of $\approx 2$ M$_{\Sun}$ yr$^{-1}$. Thus, this galaxy is a luminous S0 galaxy undergoing a starburst and not an anomalous elliptical or cD-like galaxy. Also, we do not observe a plethora of faint companions to KIG 284, as seen around NGC 1132, the isolated elliptical with detected extended X-ray emission \citep{mulchaey99}.

In summary, KIG 284 was not detected by CHANDRA at a level at least 5 times less than expected if it were the final stage in the merger history of a dense group of galaxies. But, because we have shown that this galaxy is an S0, not a luminous E, the CHANDRA non-detection is not a definitive test of the merger hypothesis for very isolated ellipticals.

\section{Conclusions and Discussion}

In this paper we have presented a sample of bright (m$_{Zw} \leq 15.7$), very isolated, early-type galaxies that can be used to investigate how early-type galaxies are formed. Starting with the KIG Catalogue of $>1000$ very isolated galaxies selected by \citet{karachentseva73} from the CGCG, we used the POSS and deeper images to scrutinize more than 200 KIG galaxies as potential Es and S0s. Table \ref{tbl:objlist} presents the basic data, including luminosities and recession velocities, for the 65 ellipticals and 37 S0 galaxies in the very isolated early-type galaxy sample that resulted from this detailed examination process. We emphasize that this sample is representative and not complete because it does not include very isolated galaxies which, by chance, are projected close to foreground or background galaxies that happen to have a similar angular size. We have used this sample to construct luminosity functions (LFs) for these galaxies and have compared them to the LFs of \citet{marzke94b} for ellipticals and S0s found throughout the CfA redshift survey region (and therefore, biased heavily towards ellipticals and S0s in rich galaxy regions). After appropriate scaling, we find that the LFs for ellipticals and S0s from \citet{marzke94b} are excellent matches to the very isolated E and S0 LFs. However, the relative numbers of Es and S0s in the \citet{marzke94b} and the KIG samples are quite different. In the CfA survey S0s outnumber Es $\sim$ 4:1 but in the KIG Catalogue (see Table \ref{tbl:objlist}), ellipticals outnumber S0s nearly 2:1.

We interpret these LF results to mean that very isolated environments are just as likely to form very luminous ellipticals as are dense groups and rich clusters. Since it is now thought that many or perhaps all elliptical galaxies formed by mergers of disk galaxies \citep{barnes92,mihos95a}, our LF results for KIG galaxies are not inconsistent with the hypothesis that very isolated Es formed by mergers as well. This is a particularly interesting suggestion for the most luminous ellipticals, whose merger histories in clusters and dense groups could be quite different from the mergers that occasionally occur in sparser environments. If the very luminous Es in the KIG sample formed through mergers, then they should be the best cases in the KIG sample for being the merger remnants of entire dense groups of galaxies.

We intended to further test the merger hypothesis for forming very isolated ellipticals specifically by searching for extended X-ray emission around a few of the most luminous KIG Es. Luminous diffuse X-ray emission would be expected if these ellipticals were the final stage in the evolution of poor, compact groups of galaxies \citep{mulchaey98}. \citet{mulchaey99} have reported the detection of one isolated elliptical, NGC 1132, at L$_x = 5 \times 10^{42}$h$^{-2}_{70}$ ergs s$^{-1}$ and \citet{mulchaey98} have detected a few elliptical-dominated poor groups of galaxies at comparable L$_x$. Thus, we had hoped to observe several luminous KIG ellipticals at or below this sensitivity limit to characterize the X-ray properties of isolated Es. However, only one of our four proposed targets was observed with CHANDRA/ACIS-S and we have now shown this galaxy, KIG 284, to be a luminous S0 and not an elliptical. Therefore, our single non-detection at L$_x < 10^{41}$h$^{-2}_{70}$ ergs s$^{-1}$, a full factor of ten below the NGC 1132 detection, does not constrain significantly the origins of the KIG isolated ellipticals or S0s. And since our upper limit on KIG 284 is comparable to the L$_x$ observed for some S0 galaxies \citep[e.g., NGC 1553;][]{blanton01}, we have no evidence that KIG 284 has abnormal X-ray properties for an S0 galaxy. Therefore, further CHANDRA observations are required to test the merger hypothesis for these systems.

While there is now significant evidence that many or all ellipticals formed by merging of disk galaxies, the origins of S0 galaxies are more obscure. Because most S0 studies have concentrated on cluster S0s, the proposed origins of these systems mostly involve processes which remove gas from disk galaxies and thus truncate star formation \citep{mihos95b,quilis00}. However, some of these processes may not be relevant to the histories of very isolated S0 galaxies. The existence of isolated S0s at luminosities comparable to group and cluster S0s suggests that there are at least two ways in which S0s can form, one method which is operable where a dense intra-group or intra-cluster gas is present and one which does not require external gas. However, the dearth of S0s compared to ellipticals (1:1.75) in the KIG sample is in great contrast to their relative abundance in the general population (S0s outnumber Es 4:1 in the CfA survey). This argues that gas stripping or other removal processes that involve a dense external medium are the most efficient method for forming S0 galaxies.

\begin{acknowledgments}
JTS, BAK, and ADL thank CHANDRA General Observer grant GO1-2090X for financial support of this work. JTS also thanks the CHANDRA grants program for providing the funds which has allowed publication of long dormant research on isolated elliptical galaxies. JTS and BAK thank the APO 3.5m observing specialists for expert assistance in obtaining some of the imaging and spectroscopy described herein.
\end{acknowledgments}

\clearpage

\clearpage
\begin{deluxetable}{lccccc}

\tabletypesize{\scriptsize}
\tablecaption{KIG Isolated Early Type Galaxies \label{tbl:objlist}}
\tablewidth{0pt}

\tablehead{\colhead{KIG \#} & \colhead{Galaxy Type} & \colhead{m$_{Zw}$} &
\colhead{cz (km/s)} & \colhead{M$_{Zw}$} & \colhead{Comments}}

\startdata
  14\tndag  & S0         & 14.7 &  5800 & $-20.1$ & \\
  19\tnddag & E          & 15.4 &  5440 & $-19.2$ & \\
  25\tnddag & E          & 14.2 &  5600 & $-20.5$ & \\
  43        & E          & 15.5 &  5280 & $-18.9$ & \\
  57        & S0         & 15.5 & 16930 & $-21.5$ & [O~II], [O~III] \\
  83\tndag  & S0         & 14.4 &  5260 & $-20.2$ & \\
  89        & S0\tnstar  & 12.6 &  1910 & $-19.7$ & \\
  91\tndag  & S0         & 14.5 &  5030 & $-20.0$ & \\
  93        & E          & 15.7 & 13730 & $-21.0$ & \\
  99        & S0         & 15.5 &  9790 & $-20.4$ & H$\alpha$ \\
 101\tndag  & S0         & 15.2 &  6090 & $-19.6$ & \\
 110\tnddag & E          & 15.0 &  6250 & $-19.9$ & \\
 111        & E\tnstar   & 15.7 &  5500 & $-19.0$ & \\
 118        & E          & 15.5 &  8200 & $-20.2$ & \\
 127        & E          & 15.7 &  7680 & $-19.7$ & \\
 128\tndag  & S0         & 15.0 &  6430 & $-20.4$ & \\
 141        & S0         & 15.0 &  1990 & $-17.9$ & [O~II], [O~III], H$\beta$ \\
 164        & E          & 15.5 &  9360 & $-20.6$ & [O~II], [O~III], H$\beta$ \\
 174\tnddag & E\tndstar  & 15.4 & 10750 & $-20.8$ & \\
 178\tnddag & E          & 15.3 &  7750 & $-20.2$ & \\
 179\tndag  & S0         & 15.3 &  5850 & $-19.7$ & [O~II] \\
 189\tnddag & E\tnstar   & 14.3 &  3150 & $-19.2$ & \\
 228        & E          & 15.6 &  8530 & $-20.0$ & \\
 245\tnddag & E\tnstar   & 15.4 &  4040 & $-18.5$ & \\
 256        & S0         & 15.5 &  6690 & $-19.5$ & \\
 264\tnddag & E          & 15.1 &  7530 & $-20.2$ & \\
 284\tndag  & S0         & 15.4 & 12990 & $-21.2$ & [O~II], H$\beta >$ [O~III]; Balmer absorption \\
 303\tndag  & S0         & 13.3 &  4060 & $-20.6$ & \\
 380        & E\tnstar   & 15.7 &  5750 & $-18.9$ & \\
 387        & E\tnstar   & 15.6 &  9060 & $-20.0$ & \\
 393\tnddag & E\tndstar  & 14.2 &  3100 & $-19.0$ & [O~II], H$\beta >$ [O~III] \\
 396\tndag  & S0         & 14.3 &  3250 & $-19.0$ & \\
 413\tndag  & S0         & 15.3 &  6320 & $-19.6$ & \\
 415        & E          & 15.5 & 11730 & $-20.7$ & \\
 417        & E\tndstar  & 15.6 &  9850 & $-20.1$ & \\
 424        & E\tnstar   & 15.6 &  7780 & $-19.6$ & Weak [O~II], H$\beta$; Balmer absorption \\
 427        & S0         & 15.5 &  7010 & $-19.5$ & \\
 430        & E          & 15.5 & 12650 & $-20.8$ & [O~II], H$\beta$ \\
 437\tnddag & E\tnstar   & 14.6 &  7430 & $-20.5$ & \\
 452        & E\tndstar  & 15.6 &  7100 & $-19.5$ & [O~II] \\
 480\tndag  & S0         & 15.2 &  5570 & $-19.3$ & [O~II], [O~III] \\
 490\tndag  & S0         & 14.8 &  5380 & $-19.7$ & \\
 501        & E          & 15.5 &  7150 & $-19.6$ & CD-like; [O~II] \\
 503        & S0         & 14.2 &  1160 & $-16.9$ & [O~II], H$\beta \gg$ [O~III] \\
 504\tndag  & S0         & 15.2 &  6020 & $-19.6$ & [O~II], H$\beta$; Balmer absorption \\
 517        & E\tnstar   & 15.5 &  9640 & $-20.2$ & \\
 555        & E\tndstar  & 15.5 &  1140 & $-15.6$ & [O~II], [O~III]; Balmer absorption \\
 556        & E          & 15.5 &  1020 & $-15.4$ & [O~II]; Balmer absorption \\
 570\tnddag & E\tndstar  & 15.4 &  5560 & $-19.2$ & \\
 574        & E\tndstar  & 15.7 & 11100 & $-20.6$ & \\
 578\tnddag & E\tndstar  & 15.2 &  9090 & $-20.4$ & \\
 582        & S0         & 15.7 & 10420 & $-20.2$ & [O~II]: \\
 589        & E\tndstar  & 15.7 & 18600 & $-21.4$ & [O~II] \\
 599\tndag  & S0\tndstar & 14.9 & 10860 & $-21.0$ & \\
 602        & S0\tndstar & 15.7 & 16180 & $-21.1$ & \\
 614        & E\tnstar   & 15.7 & 17170 & $-21.3$ & [O~II] \\
 623\tndag  & S0\tndstar & 15.4 &  5660 & $-19.2$ & \\
 636        & S0\tndstar & 15.7 & 11870 & $-20.5$ & \\
 640\tnddag & E\tndstar  & 15.1 &  9150 & $-20.5$ & \\
 670\tnddag & E          & 15.1 & 12590 & $-21.2$ & \\
 684\tnddag & E          & 14.6 &  5760 & $-20.1$ & [O~II] \\
 685        & E\tnstar   & 15.7 & 15540 & $-21.0$ & \\
 690\tndag  & S0         & 15.2 &  8020 & $-20.1$ & \\
 701        & E          & 15.6 & 24210 & $-22.1$ & \\
 703        & E\tndstar  & 15.7 &  6330 & $-19.3$ & \\
 705\tnddag & E          & 15.2 & 12160 & $-21.0$ & \\
 722\tnddag & E          & 14.9 & 10510 & $-21.1$ & \\
 732\tnddag & E          & 13.8 &  5820 & $-20.8$ & \\
 735        & E          & 15.6 &  9770 & $-20.1$ & \\
 769        & S0\tndstar & 13.5 &   310 & $-14.9$ & \\
 770\tnddag & E          & 15.1 & 12520 & $-21.3$ & \\
 771        & E          & 15.6 & 11040 & $-20.6$ & CD-like \\
 792        & E          & 15.5 &  9570 & $-20.5$ & \\
 803        & E\tnstar   & 15.5 &  8600 & $-20.0$ & \\
 811\tndag  & S0         & 15.0 &  8130 & $-20.5$ & \\
 816        & E          & 15.5 &  6880 & $-20.1$ & \\
 820        & S0         & 15.5 &  7290 & $-20.1$ & Balmer absorption \\
 823\tnddag & E          & 15.3 &  7030 & $-20.1$ & CD-like \\
 824\tnddag & E          & 14.8 &  5670 & $-19.9$ & Balmer absorption \\
 826        & E          & 15.6 &  9410 & $-20.2$ & \\
 827        & E          & 15.7 &  4600 & $-18.5$ & [O~II], H$\beta >$ [O~III] \\
 833        & E          & 15.3 &  1840 & $-17.1$ & [O~III]; Balmer absorption \\
 835\tndag  & S0\tnstar  & 15.4 &  4170 & $-18.8$ & \\
 836        & E          & 15.6 & 14900 & $-21.2$ & Radio Source \\
 841\tnddag & E\tndstar  & 14.0 &  6230 & $-20.9$ & \\
 845        & S0\tnstar  & 15.5 &  5710 & $-19.2$ & \\
 865        & S0\tnstar  & 15.6 &  7390 & $-19.7$ & \\
 877\tnddag & E\tnstar   & 15.2 &  8960 & $-20.7$ & [O~II], [O~III] $\gg$ H$\beta$; [Ne~III]  \\
 896        & S0\tnstar  & 15.7 & 10440 & $-20.5$ & \\
 898        & E          & 15.7 & 15150 & $-21.3$ & \\
 903        & S0\tnstar  & 15.5 &  5750 & $-19.3$ & [O~II] \\
 918        & E          & 15.7 &  8400 & $-20.1$ & \\
 920        & S0         & 15.6 &  5200 & $-18.9$ & \\
 921        & E\tndstar  & 15.6 &  8560 & $-20.2$ & [O~II], H$\gamma$; [O~III] $>$ H$\beta$ \\
 928        & E          & 15.7 &  7580 & $-19.7$ & [O~II], H$\beta$ \\
 981\tndag  & S0         & 15.1 &  8020 & $-20.4$ & Balmer absorption \\
1015\tnddag & E          & 14.6 &  4710 & $-19.7$ & CD-like; Balmer absorption \\
1026        & E          & 15.6 & 12780 & $-20.8$ & \\
1029        & S0         & 15.7 & 12560 & $-20.7$ & [O~II]: \\
1031        & E          & 15.5 &  1710 & $-16.7$ & [O~II], H$\gamma$, H$\beta \approx$ [O~III] \\
1042\tnddag & E\tnstar   & 14.8 & 13200 & $-21.8$ & \\
1045\tnddag & E          & 13.0 &  4020 & $-21.0$ & \\
\enddata

\tablenotetext{\dag}{~~= Member of complete sample for S0 luminosity function}
\tablenotetext{\dag\dag}{~= Member of complete sample for E luminosity function}
\tablenotetext{*}{~~= Type determined by visual inspection of new images}
\tablenotetext{**}{~= Type determined by visual inspection of POSS}

\end{deluxetable}


\clearpage
\begin{thebibliography}

\bibitem[Adams, Jensen, \& Stocke(1980)]{adams80} Adams, M., Jensen, E., \& Stocke, J. T. 1980, \aj, 85, 1010

\bibitem[Athanassoula, Makino, \& Bosma(1997)]{athanassoula97} Athanassoula, E., Makino, J., \& Bosma, A. 1997, \mnras, 286, 825

\bibitem[Barnes(1985)]{barnes85} Barnes, J. E. 1985, \mnras, 215, 517

\bibitem[Barnes \& Hernquist(1992)]{barnes92} Barnes, J. E. \& Hernquist, L. 1992, \araa, 30, 705

\bibitem[Bekki(1998)]{bekki98} Bekki, K. 1998, \apjl, 502, L133

\bibitem[Blanton, Sarazin, \& Irwin(2001)]{blanton01} Blanton, E. L., Sarazin, C. L., \& Irwin, J. A. 2001, \apj, 552, 106

\bibitem[Burstein \& Heiles(1982)]{burstein82} Burstein, D. \& Heiles, C. 1982, \aj, 87, 1165

\bibitem[Christlein \& Zabludoff(2003)]{christlein03} Christlein, D. \& Zabludoff, A. I. 2003, \apj, 591, 764

\bibitem[de Lapparent(2003)]{delapparent03} de Lapparent, V. 2003, \aap, 408,
845

\bibitem[de Vaucouleurs et al.(1976)]{devaucouleurs76} de Vaucouleurs, G., de Vaucouleurs, A., \& Corwin H. G. 1976, Second Reference Catalogue of Bright Galaxies (Austin: U. Texas Press)

\bibitem[Dressler(1984)]{dressler84} Dressler, A. 1984, \araa, 22, 1855

\bibitem[Eskridge, Fabbiano, \& Kim(1995)]{eskridge95} Eskridge, P. B., Fabbiano, G., \& Kim, D. W. 1995, \apjs, 97, 141

\bibitem[Felten(1976)]{felten76} Felten, J. E. 1976, \apj, 207, 700

\bibitem[Gaztanaga \& Dalton(2000)]{gaztanaga00} Gaztanaga, E. \& Dalton, G. B. 2000 \mnras, 312, 417

\bibitem[Haynes \& Giovanelli(1980)]{haynes80} Haynes, M. P. \& Giovanelli, R. 1980, \apjl, 240, L87

\bibitem[Haynes \& Giovanelli(1984a)]{haynes84a} Haynes, M. P. \& Giovanelli, R. 1984a, \apj, 89, 758

\bibitem[Haynes \& Giovanelli(1984b)]{haynes84b} Haynes, M. P. \& Giovanelli, R. 1984b, \aj, 89, 1

\bibitem[Haynes, Giovanelli, \& Chincarini(1984)]{haynes84c} Haynes, M. P., Giovanelli, R., \& Chincarini, G. L. 1984, \araa, 22, 445

\bibitem[Huchra(1976)]{huchra76} Huchra, J. P. 1976, \aj, 81, 552

\bibitem[Huchra et al.(1983)]{huchra83} Huchra, J. P., Davis, M., Latham, D., \& Tonry, J. 1983, \apjs, 52, 89

\bibitem[Hutchmeier \& Richter(1987)]{hutchmeier87} Hutchmeier, A. \& Richter, H. 1987, \aap, 203, 237

\bibitem[Impey \& Bothun(1997)]{impey97} Impey, C. D. \& Bothun, G. D. 1997, \araa, 35, 267

\bibitem[Karachentseva(1973)]{karachentseva73} Karachentseva, V. E. 1973, Comm. Spec. Ap. Obs. USSR, 8, 1

\bibitem[Kewley et al.(2001)]{kewley01} Kewley, L. J., Dopita, M. A., Sutherland, R. S., Heisler, C. A., \& Trevena, J. 2001, \apj, 556, 121

\bibitem[Koopmann \& Kenney(1998)]{koopmann98} Koopmann, R. \& Kenney, J. D. P. 1998, \apjl, 497, L75

\bibitem[Maccagni et al.(1987)]{maccagni87} Maccagni, D., Gioia, I. M., Maccacaro, T., Schild, R., \& Stocke, J. T. 1987, \apj, 316, 132

\bibitem[Marzke et al.(1994a)]{marzke94a} Marzke, R. O., Huchra, J. P., \& Geller, M. J. 1994a, \apj, 428, 43

\bibitem[Marzke et al.(1994b)]{marzke94b} Marzke, R. O., Huchra, J. P., Geller, M. J., \& Corwin, H. G. 1994b, \aj, 108, 437

\bibitem[Mihos(1995)]{mihos95a} Mihos, J. C. 1995, \apjl, 438, L75

\bibitem[Mihos et al.(1995)]{mihos95b} Mihos, J. C., Walker, I. R., Hernquist, L., Mendes de Oliveira, C., \& Bolte, M. 1995, \apj, 447, L87

\bibitem[McGaugh, Bothun, \& Schombert(1995)]{mcgaugh95} McGaugh, S. S., Bothun, G. D., \& Schombert, J. M. 1995, \aj, 110, 573

\bibitem[McGaugh(1996)]{mcgaugh96} McGaugh, S. S. 1996, \mnras, 280, 337

\bibitem[Mulchaey \& Zabludoff(1998)]{mulchaey98} Mulchaey, J. S. \& Zabludoff, A. I. 1998, \apj, 496, 73

\bibitem[Mulchaey \& Zabludoff(1999)]{mulchaey99} Mulchaey, J. S. \& Zabludoff, A. I. 1999, \apj, 514, 133

\bibitem[Nilson(1973)]{nilson73} Nilson, P. 1973, Uppsala General Catalog of Galaxies (Uppsala: Societis Scientiarum Upsaliensis)

\bibitem[Oemler(1992)]{oemler92} Oemler, A. 1992, in Clusters \& Superclusters, ed. A. Fabian (NATO ASI series), 29

\bibitem[Peebles(1993)]{peebles93} Peebles, P. J. E. 1993, Principles of Physical Cosmology (Princeton: Princeton U.)

\bibitem[Quilis, Moore, \& Bower(2000)]{quilis00} Quilis, V., Moore, B., \& Bower, R. 2000, Science, 288, 1617

\bibitem[Schechter(1976)]{schecter76} Schechter, P. L., 1976, \apj, 203, 297

\bibitem[Schild \& Kent(1981)]{schild81} Schild, R. \& Kent, S. 1981, SPIE, 290, 186

\bibitem[Schild(1984)]{schild84} Schild, R. 1984, \apj, 286, 932

\bibitem[Stocke(1978)]{stocke78} Stocke, J. T. 1978, \aj, 83, 348

\bibitem[Toledo et al.(1999)]{toledo99} Toledo, H. M., Dultzin-Hacyan, D., Gonzalez, J. J., \& Sulentic, J. W. 1999, \aj, 118, 108

\bibitem[Tully(1987)]{tully87} Tully, R. B. 1987, Nearby Galaxies Catalog (Cambridge: Cambridge U.)

\bibitem[Welch \& Sage(2003)]{welch03} Welch, G. L. \& Sage, L. J. 2003, \apj, 584, 260

\bibitem[Zabludoff et al.(1996)]{zabludoff96} Zabludoff, A. I., Zaritsky, D., Lin, H., Tucker, D., Hashimoto, Y., Shectman, S. A., Oemler, A., \& Kirshner, R. P. 1996, \apj, 466, 104

\bibitem[Zabludoff(2003)]{zabludoff03} Zabludoff, A. I. 2003, in The IGM/Galaxy Connection, ed. J. L. Rosenberg \& M. E. Putman (Dordrecht: Kluwer), 291

\bibitem[Zwicky et al.(1957)]{zwicky57} Zwicky, F. et al. 1957, Catalog of Galaxies \& Clusters of Galaxies, Vol. 1-6 (Pasadena: Cal Tech)

\end{thebibliography}
\end{document}